\documentclass[pdflatex,sn-mathphys-num]{sn-jnl}%

\usepackage{graphicx}%
\usepackage{multirow}%
\usepackage{amsmath,amssymb,amsfonts}%
\usepackage{amsthm}%
\usepackage{mathrsfs}%
\usepackage{physics}%
\usepackage{bm}%
\usepackage[title]{appendix}%
\usepackage{xcolor}%
\usepackage{textcomp}%
\usepackage{manyfoot}%
\usepackage{booktabs}%
\usepackage{algorithm}%
\usepackage{algorithmicx}%
\usepackage{algpseudocode}%
\usepackage{listings}%
\usepackage{setspace}
\begin{document}

\title[Article Title]{Topological mechanical neural networks as classifiers through in situ backpropagation learning}

\author{\fnm{Shuaifeng} \sur{Li}}

\author*{\fnm{Xiaoming} \sur{Mao}}\email{maox@umich.edu}

\affil{\orgdiv{Department of Physics}, \orgname{University of Michigan}, \orgaddress{\city{Ann Arbor}, \postcode{48109}, \state{Michigan}, \country{USA}}}

\abstract{Recently, a new frontier in computing has emerged with physical neural networks~(PNNs) harnessing intrinsic physical processes for learning. Here, we explore topological mechanical neural networks~(TMNNs) inspired by the quantum spin Hall effect~(QSHE) in topological metamaterials, for machine learning classification tasks. TMNNs utilize pseudospin states and the robustness of the QSHE, making them damage-tolerant for binary classification. We first demonstrate data clustering using untrained TMNNs. Then, for specific tasks, we derive an in situ backpropagation algorithm -- a two-step, local-rule method that updates TMNNs using only local information, enabling in situ physical learning. TMNNs achieve high accuracy in classifications of Iris flowers, Penguins, and Seeds while maintaining robustness against bond pruning. Furthermore, we demonstrate parallel classification via frequency-division multiplexing, assigning different tasks to distinct frequencies for enhanced efficiency. Our work introduces in situ backpropagation for wave-based mechanical neural networks and positions TMNNs as promising neuromorphic computing hardware for classification tasks.
}


\maketitle
\newpage
\section{Introduction}\label{sec1}
In the field of computational science, the quest for enhanced processing power and efficiency is never ending, especially given the recent boom in artificial intelligence. Conventional digital computing, while popular and versatile, faces imminent challenges such as energy efficiency and processing speed. One example is artificial neural networks, inspired by the intricate networks of neurons in brains, becoming a cornerstone in the advancement of machine learning and artificial intelligence~\cite{lecun2015deep}. However, training such networks on conventional von Neumann computers involves significant computational resources, energy expenditure and limitations in processing vast datasets~\cite{sze2017hardware,xu2018scaling}. This disconnection between the architecture of traditional computers and the nature of neural computation gives rise to inefficiencies in data processing. Therefore, the physical realization of analog computers is gaining attention as an alternative avenue, where the information is encoded in hardware by continuous physical variables, and where the data is processed harnessing their intrinsic ability, i.e., laws of physics~\cite{markovic2020physics}.

Physical computation embraces the complexity of physical systems. One typical example is physical neural networks~(PNNs) which represent an integration of computation form and physical systems. For example, molecular neural networks use the information-processing capabilities of biochemical sequences~\cite{bray1995protein,evans2024pattern}; optical neural networks utilize the interference patterns of light to perform machine learning, offering an unprecedented data processing speed~\cite{wetzstein2020inference}; mechanical neural networks~(MNNs) leverage deformation under applied forces to implement machine learning, resistant to the complex electromagnetic environment~\cite{stern2020supervised,li2024training}. These emerging modalities of PNNs present compelling advantages over their digital counterparts such as the potential to mimic the parallelism of the brain and achieve low-power consumption. As a result, discovering more physical processes with computational ability in physical systems becomes a promising direction.

Spin, as an intriguing concept in physics, offers a dimension in data encoding and physical computation. For example, an Ising machine can address certain types of classification tasks by reformulating them into optimization problems~\cite{laydevant2024training,scellier2017equilibrium}. By training a spin network using Equilibrium Propagation, an Ising machine can effectively classify images by finding energy minima, resembling the decision boundaries established in the feature space of the images~\cite{laydevant2024training}. The concept of spin seems to be often associated with quantum systems. However, it has been found in classical systems such as acoustics and mechanics~\cite{long2018intrinsic,shi2019observation,cheng2023backscatteringfreeedgestatesbands}. This principle has been employed to construct topological mechanical systems, an emerging field at the intersection of condensed matter physics and mechanics. These systems exhibit unique properties such as defect-immune wave propagation, attributed to underlying principles of topology~\cite{ni2023topological}. Due to the difficulty of realizing actual spin in mechanical systems, pseudospin with time-reversal symmetry is often used as an analogue to the quantum mechanical concept of spin. This includes phenomena such as pseudospin-orbit coupling, which leads to mechanical versions of QSHE~\cite{zhou2018quantum,chen2018elastic,shi2023topological}. In light of the binary nature of pseudospin states, the directional elastic waveguides where the wave propagates strictly along the interface without spreading to the bulk have been demonstrated numerically and experimentally~\cite{mousavi2015topologically,li2018observation,li2019valley,li2021topological,dorin2023uncovering,wu2023topological}. Nevertheless, the role of the pseudospin degree of freedom and binary nature of pseudospin states in performing machine learning in topological mechanical systems remains unexplored.

Therefore, in our work, we propose the concept of ``topological mechanical neural networks~(TMNNs)'' which combines topological mechanics with MNNs, and uses pseudospin states for binary classification. The topological protection endows these TMNNs superior robustness, where damages barely affect the machine learning model and waves in the inference process barely scatter into the bulk. Besides, we demonstrate  parallel classification by taking advantage of the wide bandwidth of topological states and the frequency-division multiplexing, which has the potential to improve the efficiency of machine learning~\cite{xu2022neuromorphic}.

To implement arbitrary classification tasks, the training method in mechanical systems exemplified by such TMNNs is significant to investigate. Generally, similar to training computer-based neural networks, gradient descent is the primary approach in PNNs to minimizing the loss function, necessitating efforts to obtain gradient information. Considering the connection between PNNs and the nature of neural computation, a concept known as \emph{physical learning} based on local rules has been proposed~\cite{stern2021supervised,stern2023learning}. For example, Equilibrium Propagation, a contrastive learning framework, has gained attention for its ability to train PNNs using only local information in a supervised manner~\cite{scellier2017equilibrium}. It leverages two equilibrium states—free and nudged—to approximate gradients through their comparison. While effective, this approach provides an approximate gradient of the loss function, which depends on the nudging strength. Besides, the physical learning beyond the quasistatic limit is gaining attention~\cite{stern2022physical}. Frequency Propagation introduces a harmonic signal at a specific frequency as an error signal to train PNNs, though it addresses static tasks~\cite{anisetti2024frequency}. Meanwhile, a backpropagation framework for training lattice-based metamaterials has been established, focusing on designing dispersion spectrum rather than directly training dynamic responses~\cite{chen2024intelligent}. Hence, establishing physical learning rules for wave dynamics in MNNs remain a significant challenge.

Furthermore, wave dynamics in real material systems are often complex and involve multiple resonant modes, leading to challenges in the dynamic modeling of physical systems and potentially causing a large simulation-reality gap~\cite{jiang2023metamaterial,wright2022deep}. This introduces another advantage of the TMNN as an ideal machine learning platform: they feature only topological states in the band gap, which narrows the simulation-reality gap and promotes experimental realizations.

In our work, we introduce a method of in situ backpropagation, consistent with local rules, to train TMNNs. The in situ backpropagation can be performed physically on TMNNs with only minor additional requirements for digital computers and in principle, without the need for simulation. We demonstrate that without any prior training, TMNNs can cluster data through the propagation of pseudospin waves under input forces. For given classification tasks such as Iris Flowers, Penguin and Seeds using different encoding methods, TMNNs can be trained using in situ backpropagation. TMNNs manifest as disordered networks after training but retain their topological characteristics, enabling the robust classifiers. Furthermore, we achieve efficient parallel classification in TMNNs through frequency-division multiplexing, allowing classification tasks to be performed simultaneously at different frequencies. Our work proposes a method for training TMNNs based on local rules and unveils the potential of TMNNs as classifiers for machine learning hardware.

\section{Results}\label{sec2}
\subsection{Untrained TMNNs as inherent classifiers}\label{sec2_1}
\begin{figure}[h!]
    \centering
    \includegraphics[width=1\textwidth]{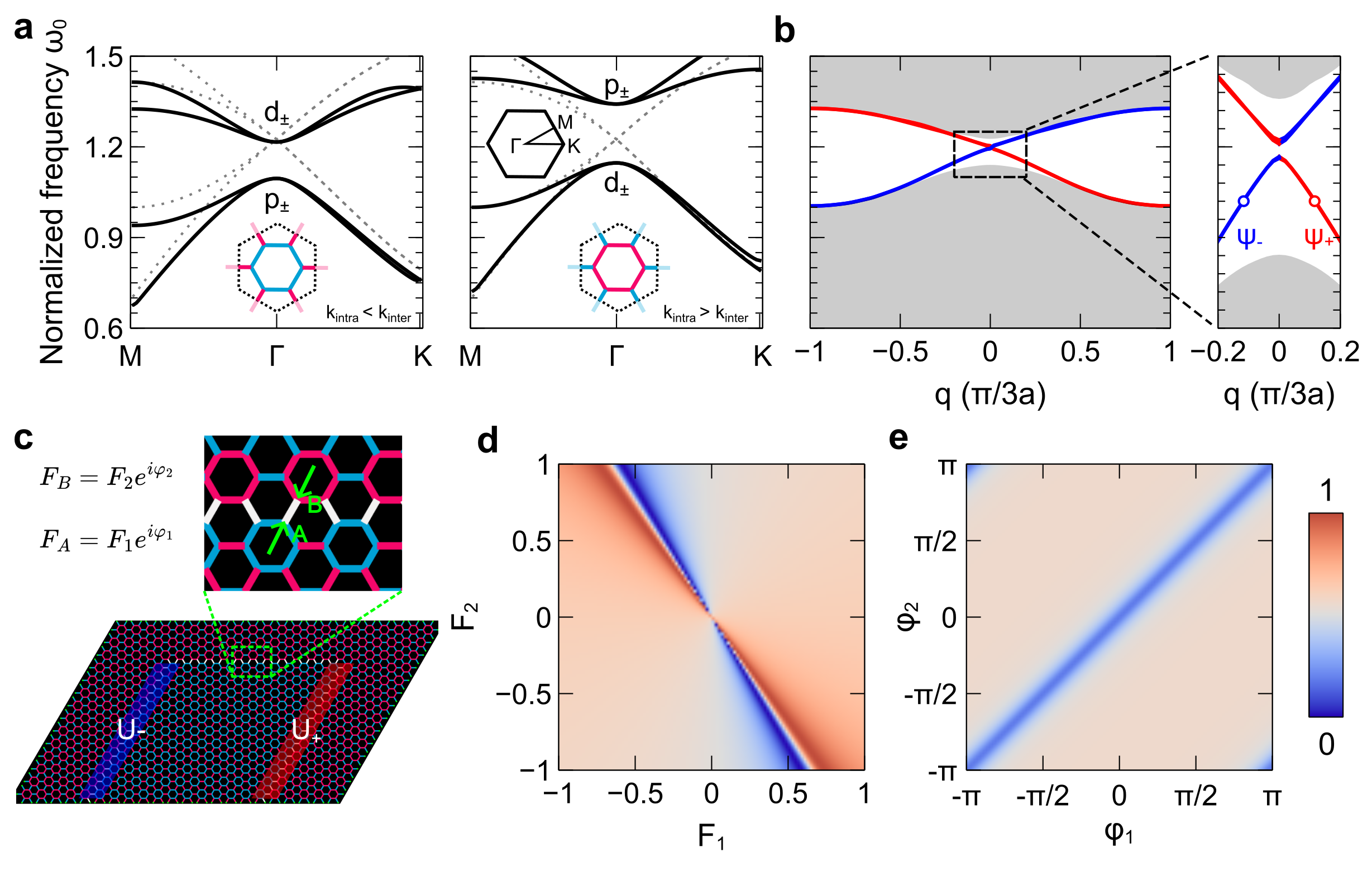}
    \caption{\textbf{Inherent capability of data clustering by untrained TMNNs.}
    \textbf{a} The first panel shows the band structure along $M$-$\Gamma$-$K$ when $\frac{k_{\mathrm{intra}}}{k_{\mathrm{inter}}}=\frac{0.8}{1.2}$ in black solid lines and the second panel shows that when $\frac{k_{\mathrm{intra}}}{k_{\mathrm{inter}}}=\frac{1.2}{0.8}$ in black solid lines. The black dashed lines represent the band structure when $\frac{k_{\mathrm{intra}}}{k_{\mathrm{inter}}}=\frac{1.0}{1.0}$.  The Brillouin zone and high symmetry points are shown in the inset. The schematics of the unit cell enclosed by black dashed lines for two cases are shown in the inset.
    \textbf{b} The projected band structure with the bulk bands~(gray shaded area) and the topological interface states~(red and blue lines). The zoom-in view is shown in the right panel. The pseudospin-up states $\Psi_{+}$ are shown in blue and the pseudospin-down states $\Psi_{-}$ are shown in red.
    \textbf{c} The TMNNs with the $U$-shaped interface. The excitations for the TMMNs are two harmonic forces with amplitudes $F_{1}$ and $F_{2}$, and phases $\varphi_{1}$ and $\varphi_{2}$. The directions of two forces are along the bond, as shown in the zoom-in view of TMNNs.
    \textbf{d} The ratio $r$ as a function of $F_{1}$ and $F_{2}$ with $\varphi_{1}=\varphi_{2}=0~\mathrm{rad}$.
    \textbf{e} The ratio $r$ as a function of $\varphi_{1}$ and $\varphi_{2}$ with $F_{1}=0.5~\mathrm{N}$, $F_{2}=-1~\mathrm{N}$.
    }
    \label{fig1}
\end{figure}

We begin by introducing TMNNs that possesses the QSHE to enable binary classification based on pseudospin waves propagation. Our TMNN is composed of two domains with different topological phases, based on the honeycomb-lattice topological mechanical metamaterial~\cite{chen2019topological}. As illustrated in the inset of Fig.~\ref{fig1}a, the unit cell of each domain enclosed by black dashed lines is composed of six nodes, six intra bonds with spring constant $k_{\mathrm{intra}}$ connecting the six nodes, and six connections to adjacent unit cells with spring constant $k_{\mathrm{inter}}$. Note that each connection is equivalent to a half inter bond shown in the dark color. When $k_{\mathrm{intra}}=k_{\mathrm{inter}}$, a double Dirac cone appears in the band structure~(depicted by black dashed lines in Fig.~\ref{fig1}a) resulting from the folding mechanism of the Brillouin zone~\cite{wu2015scheme}. One can expect that when $k_{\mathrm{intra}}$ and $k_{\mathrm{inter}}$ are unequal, a band gap will formed due to the broken symmetry, as shown in two panels of Fig.~\ref{fig1}a. As indicated by studies before, topological properties of the band gap are distinct depending on the values of $k_{\mathrm{intra}}$ and $k_{\mathrm{inter}}$~\cite{zhou2018quantum,shi2023topological}. In the first panel of Fig.~\ref{fig1}a, when $k_{\mathrm{intra}}<k_{\mathrm{inter}}$, this domain is considered as the topologically nontrivial phase, since the $d_{\pm}$ states dominate the lower bands while the $p_{\pm}$ states dominate the higher bands. In contrast, as shown in the second panel of Fig.~\ref{fig1}a, when $k_{\mathrm{intra}}>k_{\mathrm{inter}}$, this domain exhibit the topologically trivial phase, where $d_{\pm}$ and $p_{\pm}$ states invert.

When two domains with opposite topological phases are joint together, the projected band structure along $\Gamma K$ direction is characterized by two bands representing topological states in the bulk band gap, as presented in Fig.~\ref{fig1}b. The zoom-in view is shown in the right panel. These two bands also represent two different pseudospin states -- pseudospin-up state $\Psi_{+}$ and pseudospin-down state $\Psi_{-}$ -- with opposite propagation directions in group velocity $v=\frac{\partial\omega}{\partial{k}}$. Besides, topological states propagate and are well-confined along the interface. This motivates us to use pseudospin waves for binary classification in machine learning by leveraging distinct propagation directions.

For demonstration, in Fig.~\ref{fig1}c, we construct a two-dimensional TMNN formed by the nontrivial domain as inner network and the trivial domain as outer network, resulting in a $U$-shaped interface. We define our TMNN as a entire network with $n$ nodes located at position $\{x_{p}\}$, connected by $l$ linear, non-dissipative springs, each with a spring constant $k_{i}$. The connectivity of the TMNN is described by the compatibility matrix $C\in\mathbb{R}^{l\times 2n}$, which maps the $2n$ displacement of the $n$ nodes in two dimensions to the $l$ bond elongations in the linear regime following $E=CU$~\cite{pellegrino1986matrix,sun2012surface}. This relation is expressed as $E_{i}=\hat{I}_{j_{1}j_{2}}\cdot (U_{j_{1}}-U_{j_{2}})$, where $\hat{I}_{j_{1}j_{2}}$ is a unit vector pointing from node $j_{1}$ to node $j_{2}$ determining each entry in $C$. Note that in the frequency domain we use the capital letter $U$ and $E$ to represent modal displacement and modal elongation to differ from the case in the time domain. Besides, the dynamical matrix $D$ of the TMNN can be described as $D=C^{T}KC$, where $K$ is the diagonal matrix with spring constants of bonds as the diagonal entries such that $K_{ii}=k_{i}$. 

Previous studies have indicated that pseudospin-dependent states can be selectively excited in the topological mechanical systems~\cite{chaunsali2018experimental,chaunsali2018subwavelength,wu2018dial,chen2019topological}, due to the effective model of interface states under the basis of pseudospin-down state~($p_{+}$ and $d_{+}$) and pseudospin-up state~($p_{-}$ and $d_{-}$)~\cite{chen2019topological}:
\begin{equation}
    \label{equ1}
    \psi_{\pm}(x,y)=\frac{1}{\sqrt{2}}\begin{bmatrix}
    -1\\e^{\pm i\theta}
    \end{bmatrix}e^{-|\frac{\delta t}{v}x|}e^{i\Delta k_{y}y},
\end{equation}
where $\theta$ is the interface angle, $\delta t=k_{\mathrm{intra}}-k_{\mathrm{inter}}$, $\Delta{k_{y}}$ is the wave vector and $x$, $y$ are spatial coordinates in the real space. When the excitation aligns more closely with $\psi_{-}$, it predominantly excites pseudospin-down states, whereas alignment with $\psi_{+}$ leads to the excitation of more pseudospin-up states. We study the excitation of pseudpspin states by using force excitation on two nodes of a bond across the interface (Fig.~\ref{fig1}c). In a two dimensional system, each node bears forces along two orthogonal directions, i.e., $F_{x}=|F_{x}|e^{i\varphi_{x}}$ and $F_{y}=|F_{y}|e^{i\varphi_{y}}$. Here we discuss in the frequency domain so time evolution of all dynamic quantities carry the same factor $e^{i\omega t}$, where $\omega$ is the angular frequency within the band gap, and can be omitted in the analysis. For the inner networks, $\frac{k_{\mathrm{intra}}}{k_{\mathrm{inter}}}=\frac{0.8}{1.2}$, and for the outer networks, $\frac{k_{\mathrm{intra}}}{k_{\mathrm{inter}}}=\frac{1.2}{0.8}$. The spring constant for the interface region $k_{\mathrm{interface}}$ remain unchanged. The mass of the nodes $m$ is unity. To generalize the description of TMNNs, we use the normalized angular frequency $\omega_{0}=\omega\sqrt{\frac{m}{k_{\mathrm{interface}}}}$ in the following discussion. 

In our demonstration of the inherent data clustering by the untrained TMNN, we fix the directions of two forces $F_{A}=F_{1}e^{i\varphi_{1}}$ and $F_{B}=F_{2}e^{i\varphi_{2}}$ to be along the bond, as depicted in Fig.~\ref{fig1}c. To check which pseudospin state is dominant, we select nodes along the interface on both sides, illustrated in Fig.~\ref{fig1}c by blue and red regions, corresponding to the modal displacements $U_{-}\in\mathbb{C}^{2n_{\mathrm{output}}\times{1}}$ and $U_{+}\in\mathbb{C}^{2n_{\mathrm{output}}\times{1}}$, respectively, to contain both horizontal and vertical displacements. The blue and red regions also indicate the propagation of pseudospin-up states and pseudospin-down states, respectively. We define the ratio between the amplitude of excited pseudospin-up states, which predominantly propagates to the $U_{+}$ region, and the total amplitude of excited states represented by the modal displacement, as follows: 
\begin{equation}
    \label{equ2}
    r=\frac{\langle U_{+}|U_{+} \rangle}{\langle U_{+}|U_{+} \rangle + \langle U_{-}|U_{-} \rangle}.
\end{equation}
This implies that when $r>0.5$, the excited states prefer to propagate rightwards, and they prefer to propagate leftwards when $r<0.5$.

In Fig.~\ref{fig1}d, we display $r$ as a function of $F_{1}$ and $F_{2}$ when $\varphi_{1}=\varphi_{2}=0~\mathrm{rad}$. When both $F_{1}$ and $F_{2}$ are in the reddish region, more excited states are pseudospin-up states and propagate rightwards ($r>0.5$). On the other band, when both $F_{1}$ and $F_{2}$ are in the blueish region, the majority of the excited states are pseudospin-down states, thus propagating leftwards ($r<0.5$). Likewise, when we fix the amplitudes of forces to be $F_{1}=0.5~\mathrm{N}$ and $F_{2}=-1~\mathrm{N}$ but vary the phases $\varphi_{1}$ and $\varphi_{2}$, $r$ as a function of $\varphi_{1}$ and $\varphi_{2}$ is illustrated in Fig.~\ref{fig1}e. By demonstrating such excitation to generate pseudospin states, we highlight a critical observation: if regarding the reddish and bluish regions as two distinct classes, we leverage TMNNs to categorize the data within the range $[-1,1]$ using amplitude encoding and within $[-\pi,\pi]$ using phase encoding, underscoring the computation and data clustering capability of TMNNs. Note that the results can vary under different configurations and boundary conditions, as the pseudospin states in a finite system differ from those in a supercell that is infinite in one direction.

However, in the context of supervised learning, the current configuration, which can only produces certain classification pattern in Figs.~\ref{fig1}d and e, is inadequate for performing arbitrary classification tasks. To address this limitation, we introduce the method of in situ backpropagation to train TMNNs below, which can be conducted locally without a centralized processor.

\subsection{In situ backpropagation in TMNNs}\label{sec2_2}
In our previous work, we proposed the in situ backpropagation for MNNs in static case~\cite{li2024training}. We now introduce the theoretical framework to conduct in situ backpropagation in TMNNs within the frequency domain, i.e., harmonic wave analysis. This involves obtaining the gradient of a loss function with respect to spring constants. This training protocol can be seamlessly applied to general elastic wave propagation in MNNs within the frequency domain, where the angular frequency $\omega$ is a given parameter. For a given classification task, the learning problem can be described as:
\begin{equation}
    \label{equ3}
    \begin{aligned}
        & \underset{k}{\text{minimize}} & & \mathcal{L}\left[U(k),U^{\dagger}(k)\right],\\
        & \text{subject to} & & (-\omega^{2}M+D)U=F,
    \end{aligned}
\end{equation}
where $\mathcal{L}$ is the loss function measuring the difference between the output and the desired output, $k\in\mathbb{R}_{\geq0}^{l\times 1}$ is a vector containing the spring constant of each bond, which is the trainable learning degree of freedom, $M\in\mathbb{R}_{>0}^{2n\times 2n}$ is the diagonal mass matrix, $\omega$ is the angular frequency of the excitation. $D\in\mathbb{R}^{2n\times 2n}$ is the symmetric stiffness matrix, $U\in\mathbb{C}^{2n\times1}$ is the modal displacement of the node, which is the output and $F\in\mathbb{C}^{2n\times1}$ is the harmonic forces applied on the nodes, nonzero values of which are the input containing both the amplitude and the phase~(e.g., $F_{A}$ and $F_{B}$ in Section~\ref{sec2_1} are entries in the vector $F$). The governing equation of harmonic wave propagation in the frequency domain, $(-\omega^{2}M+D)U=F$, represents the forward problem, reflecting the response $U$~(modal displacement of each node) under input harmonic forces $F$. To minimize $\mathcal{L}\left[U(k),U^{\dagger}(k)\right]$ using gradient descent, $\nabla\mathcal{L}$ is derived as below:
\begin{equation}
    \label{equ4}
    \nabla\mathcal{L}=\frac{d\mathcal{L}}{dk}=\frac{\partial\mathcal{L}}{\partial U}\frac{dU}{dk}+\frac{\partial\mathcal{L}}{\partial U^{\dagger}}\frac{dU^{\dagger}}{dk}=2\Re{\frac{\partial\mathcal{L}}{\partial U}\frac{dU}{dk}}.
\end{equation}
Given that the loss $\mathcal{L}$ is defined as a function of $U$, the Jacobian $\frac{\partial\mathcal{L}}{\partial U}$ can be conveniently calculated, whereas usually $\frac{dU}{dk}$ is computationally-heavy due to the complex interaction in the network. By differentiating $(-\omega^{2}M+D)U=F$ on both sides, $\frac{dU}{dk}$ can be derived:
\begin{equation}
    \label{equ5}
    \frac{dU}{dk}=-(-\omega^{2}M+D)^{-1}\frac{dD}{dk}U,
\end{equation}
with $\frac{d}{dk}F=\bm{0}$. Then plug Eq.~\eqref{equ5} to Eq.~\eqref{equ4}:
\begin{equation}
    \label{equ6}
    \begin{split}
    \nabla\mathcal{L}&=2\Re{\frac{\partial \mathcal{L}}{\partial U}\left[-(-\omega^{2}M+D)^{-1}\frac{dD}{dk}U\right]}\\
    &=2\Re{U_{\mathrm{adj}}^{T}\frac{dD}{dk}U}.
    \end{split}
\end{equation}
Here we define the transpose of the adjoint displacement $U_\mathrm{adj}^{T}$ as $-\frac{\partial \mathcal{L}}{\partial U}(-\omega^{2}M+D)^{-1}$ to obtain $U_\mathrm{adj}^{T}(-\omega^{2}M+D)=-\left(\frac{\partial \mathcal{L}}{\partial u}\right)$. Note that both $M$ and $D$ are symmetric so $(-\omega^{2}M+D)$ is symmetric. Therefore, after taking the transpose on both sides, the adjoint problem can be defined as below:
\begin{equation}
    \label{equ7}
    (-\omega^{2}M+D)U_{\mathrm{adj}}=-\left(\frac{\partial \mathcal{L}}{\partial U}\right)^{T}.
\end{equation}
Apparently, the two problems~(forward and adjoint) differ in their harmonic forces. Physically, the adjoint problem can be understood as the response $U_{\mathrm{adj}}$ under the adjoint force $-\left(\frac{\partial\mathcal{L}}{\partial U}\right)^{T}$.

The gradient of $\mathcal{L}$ in Eq.~\eqref{equ6} can be further expressed as:
\begin{equation}
    \label{equ8}
    \begin{split}    \nabla\mathcal{L}&=2\Re{U_{\mathrm{adj}}^{T}\frac{dD}{dk}U}=2\Re{U_{\mathrm{adj}}^{T}\frac{d(C^{T}KC)}{dk}U}\\
    &=2\Re{U_{\mathrm{adj}}^{T}C^{T}\frac{dK}{dk}CU}=2\Re{E_{\mathrm{adj}}\circ E},
    \end{split}
\end{equation}
where $\circ$ is the Hadamard product~(i.e., element-wise product). $\frac{dK}{dk}$ is a tensor such that $\frac{dK_{op}}{dk_{q}}=\delta_{oq}\delta_{pq}\in\mathbb{R}^{l\times l\times l}$, where the entry is $1$ when $o=p=q$ and otherwise $0$. Eq.~\eqref{equ8} implies that the gradient of the loss function $\mathcal{L}$ equals to the element-wise multiplication of modal elongations of bonds in the forward problem $(-\omega^{2}M+D)U=F$ and the adjoint problem $(-\omega^{2}M+D)U_\mathrm{adj}=-\left(\frac{\partial \mathcal{L}}{\partial U}\right)^{T}$.

This method of in situ backpropagation aligns with the ``local rule'' required in physical learning, as the gradient for bond $i$ can be obtained solely from the modal elongation of bond $i$, i.e., $\nabla\mathcal{L}_{i}=E_{\mathrm{adj},i}E_{i}$. Our in situ backpropagation method for training TMNNs in wave dynamics showcases the potential of leveraging local rules without extensive reliance on digital computers, thereby enabling high efficiency. Next, the gradient $\nabla\mathcal{L}_{i}$, obtained locally at each bond $i$ via the two steps described above, is used to update the spring constants at learning rate $\alpha$, from $k_i$ to $k_{i} - \alpha \nabla\mathcal{L}_{i}$
\begin{equation}
    \label{equ9}
    k_{i} \leftarrow k_{i} - \alpha\nabla\mathcal{L}_{i}
    = k_{i} - 2\alpha\Re{E_{\mathrm{adj}, i}E_{i}},
\end{equation}
iteratively through gradient descent, minimizing the loss function subject to the physics law.

\begin{figure}[h!]
    \centering
    \includegraphics[width=0.8\textwidth]{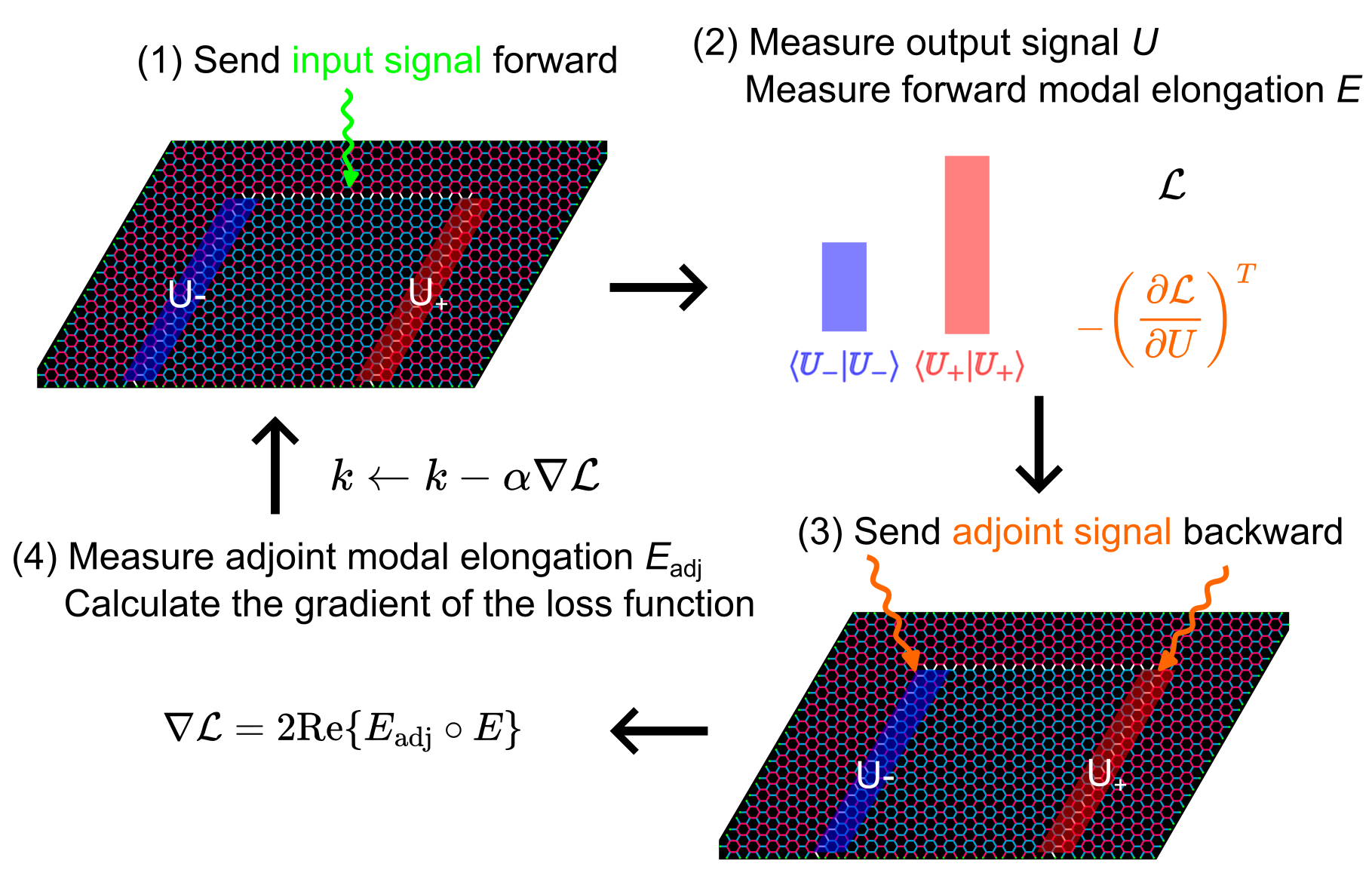}
    \caption{\textbf{In situ backpropagation to train TMNNs.}
    The schematic shows the iterative four-step procedure to obtain the gradient of the loss function locally to update spring constants of TMNNs.
    }
    \label{fig2}
\end{figure}

We summarize the four steps to implement in situ backpropagation in TMNNs, which can be done either  physically or computationally, in Fig.~\ref{fig2}. First, the untrained TMNN are initialized and the input force through various encoding methods is sent into the untrained TMNN, which essentially serves as a forward pass. Second, we measure the output modal displacement of nodes in two shaded areas, named $U_{+}$ and $U_{-}$, and subsequently plug the measured displacement into the loss $\mathcal{L}$ and into $-\left(\frac{\partial\mathcal{L}}{\partial U}\right)^{T}$ to obtain the value of adjoint force. Note that the entries of the adjoint force are nonzero only at output nodes. Simultaneously, the modal elongation $E$ of bonds is measured. Third, the adjoint force is applied into the TMNN, serving as a backward pass. Fourth, we measure the adjoint modal elongation $E_{\mathrm{adj}}$, and calculate the gradient of the loss function according to the learning rule in Eq.~\eqref{equ8}. These steps are done under the designated frequency $\omega$. Finally, spring constants in the TMNN are updated according to gradient descent in Eq.~\eqref{equ9}. Among these four steps, the calculations for the adjoint force $-\left(\frac{\partial\mathcal{L}}{\partial U}\right)^{T}$ and the gradient require digital computers. Note that the adjoint force is sparse~($\|-\left(\frac{\partial\mathcal{L}}{\partial U}\right)^{T}\|_{0}=\frac{2n_{\mathrm{output}}}{n}$, $\|\cdot\|_{0}$ is the $l_{0}$-norm measuring the number of nonzero entries) which entails minimal computational cost. The remaining steps could be done physically through the response of the TMNN.

In addition, there exists another learning degrees of freedom, the mass of the nodes $m_{j}$. Following the similar derivation based on the adjoint method, we show the learning rule for training $m$ and the update of mass, for the defined loss function $\mathcal{L}[U(m),U^{\dagger}(m)]$:
\begin{equation}
    \label{equ10}
    \nabla\mathcal{L}=\frac{d\mathcal{L}}{dm}=-2\omega^{2}\Re{\sum_{s=1}^{s=2}U_{\mathrm{adj}}^{(s)}\circ U^{(s)}},
\end{equation}
\begin{equation}
    \label{equ11}
    m_{j} \leftarrow m_{j} - \alpha \nabla\mathcal{L}_{j}
    = m_{j} + 2\alpha\omega^{2}\Re{\sum_{s=1}^{s=2}U_{\mathrm{adj}, j}^{(s)}U_{j}^{(s)}}.
\end{equation}
The detailed derivation can be seen in the Supplementary Information. This learning rule for training the mass of the nodes also features a four-step procedure shown in Fig.~\ref{fig2}, and the local rule by only knowing the modal displacement of each node.

\subsection{Training TMNNs for classification tasks}\label{sec2_3}
\begin{figure}[h!]
    \centering
    \includegraphics[width=1\textwidth]{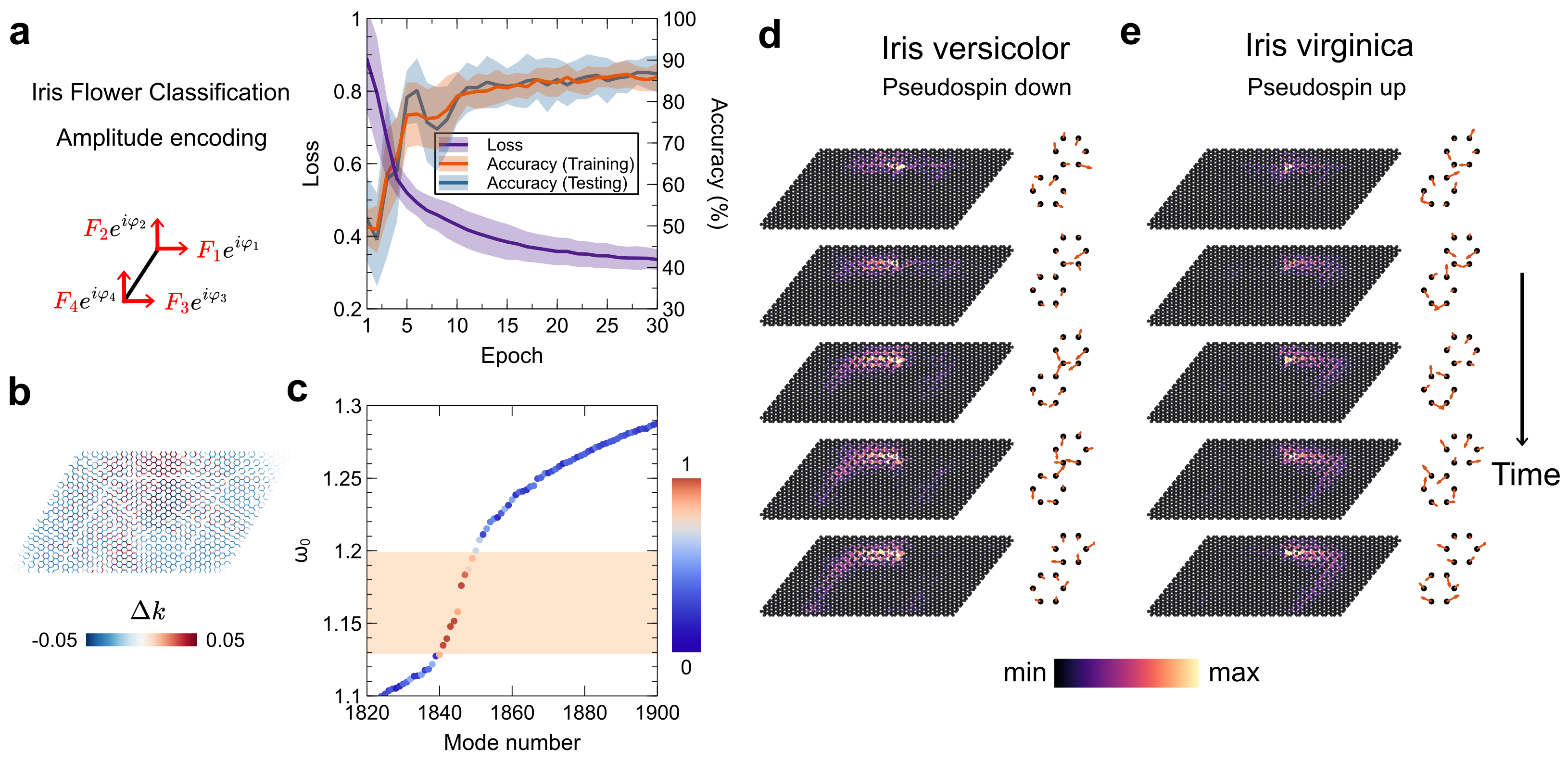}
    \caption{\textbf{Demonstrations of Iris flower classification tasks with amplitude encoding}
    \textbf{a} The loss~(purple) and classification accuracy~(orange for training set and blue for testing set) as a function of epoch in training processes of Iris flower classification under the amplitude encoding.
    \textbf{b} The difference of spring constants $\Delta{k}$ between the trained TMNNs and the untrained TMNNs.
    \textbf{c} The normalized eigenfrequency $\omega_{0}$ of the trained TMNNs. The modes in shaded regions represent the topological states whose localization is encoded by color.
    \textbf{d} \textbf{e} The inference processes in TMNNs. The wave propagation over time is shown when the input is from amplitude encoding of the features of Iris versicolor and Iris virginica. The velocity of nodes along the wave propagation direction is displayed beside the corresponding snapshots to represent the pseudospin states.
    }
    \label{fig3}
\end{figure}

For binary classification tasks, the commonly used loss function is the cross entropy loss defined as $\mathcal{L}=-[y\ln{r}+(1-y)\ln{(1-r)}]$, where $y$ is the binary indicator~($0$ or $1$). Physically, loss is minimized when the excited waves preferentially propagate in only one direction. In Section~\ref{sec2_1}, we show the potential encoding methods using amplitude and phase of forces to represent data. In the following, we employ three well-known classification tasks -- Iris flower, Penguin and Seeds -- to exemplify the classification processes using amplitude encoding, phase encoding and hybrid encoding, respectively, as shown in the Fig.~\ref{fig3}~(amplitude encoding) and Supplementary Information~(phase encoding and hybrid encoding). Note that the encoding methods for tasks are not specifically chosen, allowing for any encoding method to be used for tasks.

For Iris flower classification, the goal is to classify two types of Iris flowers -- Iris versicolor and Iris virginica -- using four distinct features: sepal length, sepal width, petal length, and petal width. In amplitude encoding, each feature corresponds to the amplitude of a harmonic force~(with phase being zero) applied to the nodes at the same time at a given $\omega$, marked in Fig.~\ref{fig1}c. These four values are scaled into the range from $-1$ to $1$, and each node can bear two harmonic forces along $x$ and $y$ directions, respectively. The Iris versicolor is classified when the excited waves propagate leftwards and the Iris virginica is classified when the excited waves propagate rightwards. The dataset is randomly partitioned into a training set~($70\%$) and a testing set~($30\%$). In Fig.~\ref{fig3}a, as the loss steadily decreases over epoch, the classification accuracy for the training dataset approaches $85\%$ on average, while the accuracy for the testing dataset, which is unseen during the training process, converges to $86\%$ on average, suggesting effective learning by TMNNs.

The difference between the trained TMNN and the untrained TMNN is exhibited in Fig.~\ref{fig3}b. Generally, the spring constants decrease for most bonds.  Importantly, to enable TMNNs to achieve such complex task, the spring constants only require less than $5\%$ change, which can be convenient for the experimental realization. We also observe that the untrained TMNN manifests as a periodic network, whereas the trained TMNN becomes a disordered network with varying spring constant distributions. This raises a natural question: does this system remain topological such that the topological edge states and pseudospin nature that the classification relies on can still exist?

Previous works in photonic, acoustic, and mechanical systems have demonstrated that topological edge states can exist even in the absence of lattice periodicity~\cite{liu2017disorder,zhou2018,zhou2019,zhang2022disorder,liu2023acoustic,shi2023topological}. While local disorder can be seen as breaking the translational periodicity in the underlying lattice, topological edge states are protected against weak disorder. However, under sufficiently strong disorder, this topological protection can break down.

To check the existence of the topological states in the band gap, we fist calculate the eigenfrequencies of the trained TMNN. As shown in Fig.~\ref{fig3}c, there are eigenmodes within the bulk band gap region for the trained TMNN. Besides, we calculate the localization of these eigenmodes represented by the weights of eigenmodes along the interface, $W_{j\in{interface}}=\frac{\langle U_{j}|U_{j}\rangle}{\langle U|U\rangle}$, which is encoded by color in Fig.~\ref{fig3}c. The larger weights indicate the stronger localization along the interface. This means that these modes in the band gap remain localized at the interface.

We then adapt the spin Bott index to characterize the topology of the trained TMNN to ensure that these localized modes are topological modes, which serves a similar function to the spin Chern number but can be applied in real space when momentum space is not well-defined~\cite{huang2018quantum,huang2018theory}. Following the calculation detailed in the Supplementary Information, the spin Bott index exhibits a nontrivial value of $1$ for the inner networks and $0$ for the outer networks. The above analysis indicates that the trained TMNN remains topological, preserving topological states and pseudospin nature, which protects the classification mechanism in TMNNs.

Finally, we demonstrate the inference process, as shown in Figs.~\ref{fig3}d and e. In Fig.~\ref{fig3}d, when the input harmonic forces are encoded from the features of Iris versicolor~(i.e., the amplitudes of the forces correspond to the features of Iris versicolor and the phases are set to zero), the excited waves propagate leftwards over time along the interface. In sharp contrast, when the input harmonic forces are encoded from the features of Iris virginica, in Fig.~\ref{fig3}e, the excited waves propagate rightwards over time along the interface. Notably, when the waves travel rightwards, they can smoothly propagate through the sharp $60^{\circ}$ bend, demonstrating the bend-immune nature of the trained TMNN with a nontrivial spin Bott index. In addition, we showcase the pseudospin states represented by the velocity of the node near the interface in the inference process. The pseudospin-down state is excited to propagate leftwards for Iris versicolor, whereas the pseudospin-up state is excited to propagate rightwards for Iris virginica.

We also showcase the phase encoding for Penguin classification task and hybrid encoding for Seeds classification task in the Supplementary Information, where, on average, $88\%$ training accuracy and $92\%$ testing accuracy for Penguin classification, and $96\%$ training accuracy and $94\%$ testing accuracy for Seeds classification are achieved. The successful classification demonstrated in three examples confirms the effective classification function of our trained TMNN. It is worthwhile to note that the elastic waves propagate along the interface smoothly even in the trained~(disordered) networks due to the topological protection, serving as a clear indicator for classification results, which is difficult to realize in conventional networks.

Furthermore, our method of in situ backpropagation can train the mass of the node instead of the spring constant to achieve classification. In the Supplementary Information, without loss of generality, we demonstrate the Iris flower classification by training the mass of the node using amplitude encoding, where the training accuracy reaches $90\%$ and the testing accuracy reaches $92\%$, with the decrease of the loss.

\subsection{Robustness of TMNNs as classifiers}\label{sec2_4}
After being trained, TMNNs remain topological states according to the analysis in the last section. However, as indicated by several numerical and experimental studies, pseudospin waves are not always fully protected~\cite{deng2017observation,chaunsali2018subwavelength}. Different types of defects have varying impacts on wave propagation. This phenomenon can be more pronounced in the trained TMNNs. There are two types of defects according to previous studies: non-spin-mixing defects that preserve the pseudospin states and spin-mixing defects that break the pseudo-time-reversal symmetry and mix the two pseudospin states. The pseudospin of the state will be flipped when encountering spin-mixing defects, while preserved under non-spin-mixing defects. Therefore, our classification mechanism based on pseudospin states can be affected by these network imperfections.

\begin{figure}[h!]
    \centering
    \includegraphics[width=0.8\textwidth]{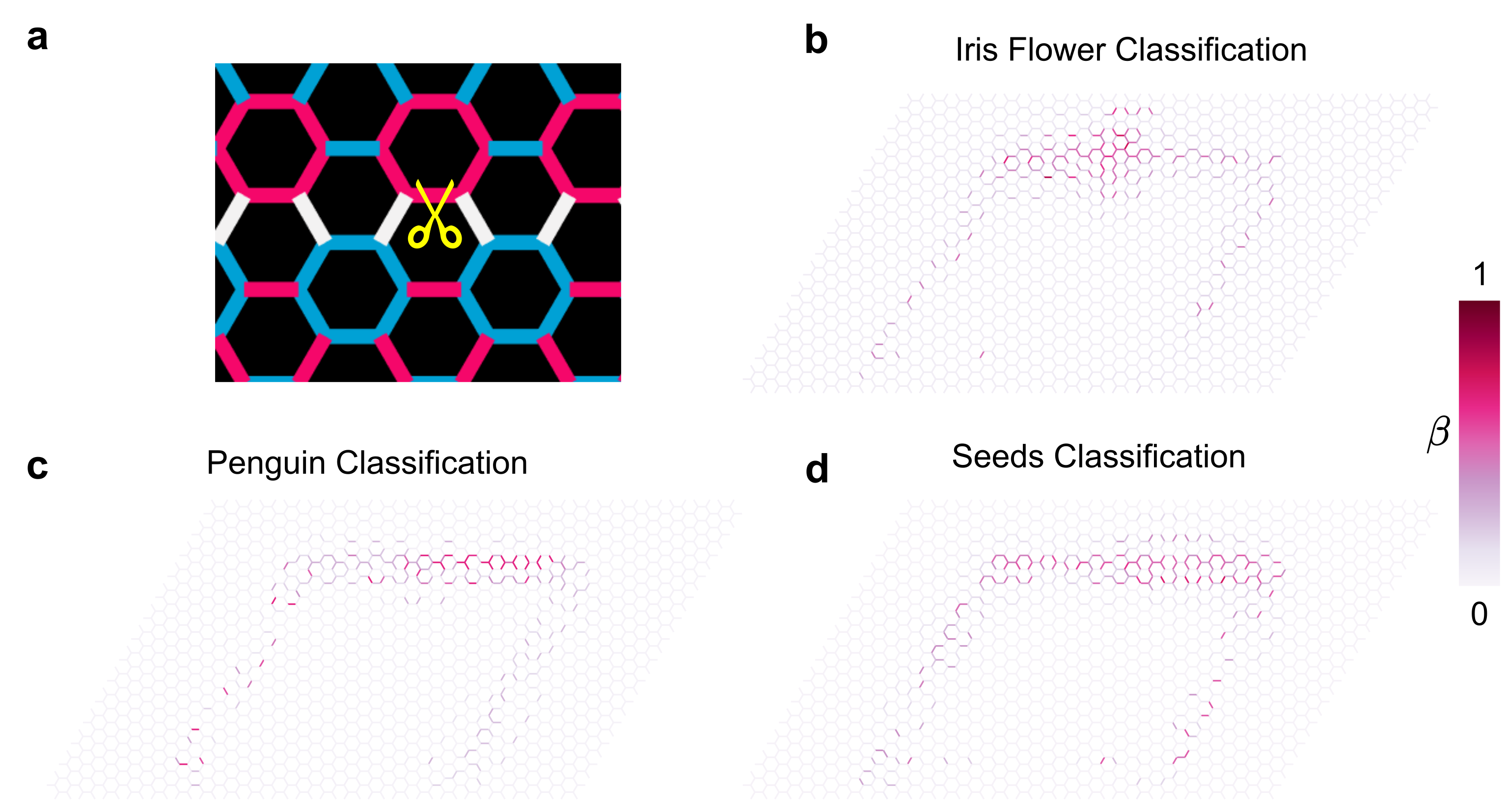}
    \caption{\textbf{Robustness of TMNNs.}
    \textbf{a} The schematics of testing robustness of TMNNs through pruning bonds.
    \textbf{b}, \textbf{c} and \textbf{d} The decrease of the classification accuracy normalized by the original classification accuracy after pruning the corresponding bond for Iris flower classification, Penguin classification and Seed classification tasks, respectively.
    }
    \label{fig4}
\end{figure}

To gauge the effect of the imperfections in each classification model discussed above, we introduce an index $\beta$ defined as the accuracy drop normalized by the original accuracy after pruning one bond in the trained TMNN shown by schematics in Fig.~\ref{fig4}a. Thus, $\beta>{0}$ indicates that the defect caused by pruning the bond is spin-mixing defect, which affects the classification model significantly. In contrary, $\beta=0$ means non-spin-mixing defects which barely affect the classification model. 

As illustrated in Figs.~\ref{fig4}b, c and d, $\beta$ for each bond is color-encoded for the classifications of Iris flower, Penguin, and Seeds, respectively. Most bonds are colored in light shades, indicating the robustness of the TMNN after pruning one of these bonds. Light-shaded bonds also represent non-spin-mixing defects, which barely affect the classification model. However, pruning bonds colored in dark shades leads to a significant drop in classification accuracy for all three cases, indicating spin-mixing defects. Besides, most light-shaded bonds are distributed in the bulk rather than at the interface, whereas most heavy-shaded bonds are located along the interface, meaning the interfacial bonds are more sensitive in the classification model. Overall, the predominance of light-shaded bonds demonstrates the robustness of TMNNs, enabling robust machine learning hardware through TMNNs. Identifying spin-mixing and non-spin-mixing defects helps in understanding the robustness and in recognizing their vulnerabilities of TMNNs based on pseudospin states .

\subsection{Parallel classification in TMNNs}\label{sec2_5}
\begin{figure}[h!]
    \centering
    \includegraphics[width=1\textwidth]{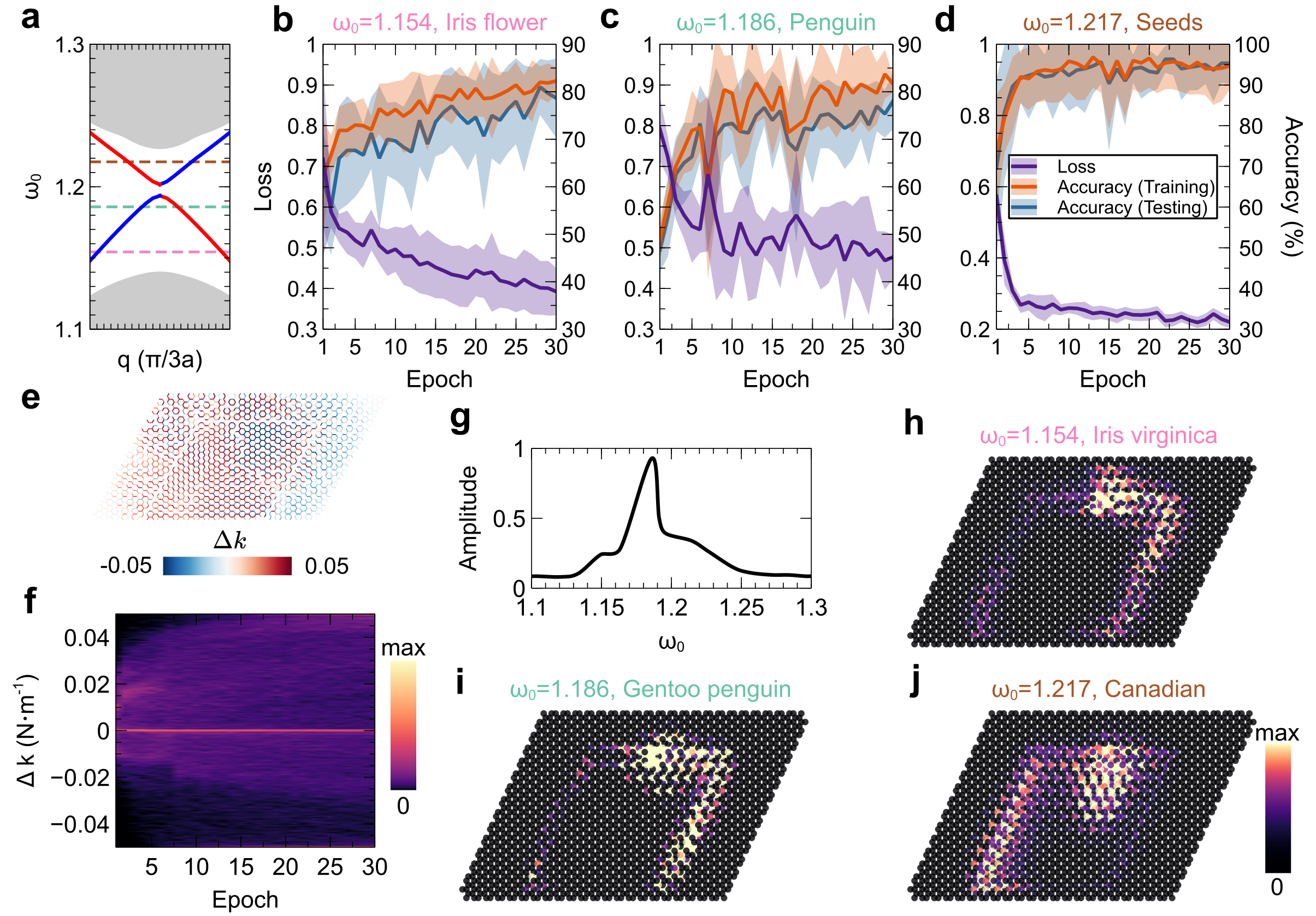}
    \caption{\textbf{Parallel classification using TMNNs.}
    \textbf{a} The projected structure with the bulk bands~(gray shaded area) and the topological interface states~(red and blue lines). Dashed lines indicate the frequencies used for frequency-division multiplexing in parallel classification.
    \textbf{b}, \textbf{c} and \textbf{d} The loss~(purple) and classification accuracy~(orange for training set and blue for testing set) as a function of epoch in training processes of parallel classification are shown separately for Iris flower, Penguin and Seeds classifications, respectively.
    \textbf{e} The difference of spring constants $\Delta{k}$ between the trained TMNNs and the untrained TMNNs.
    \textbf{f} The histogram of the difference of spring constants between the TMNNs and the untrained TMNNs $\Delta{k}$ as a function of the training epoch.
    \textbf{g} DMD spectrum shows that DMD mode amplitude varies as a function of frequency.
    \textbf{h}, \textbf{i} and \textbf{j} Amplitude of the DMD mode corresponding to $\omega_{0}=1.154$, $1.186$ and $1.217$, respectively.
    }
    \label{fig5}
\end{figure}

In the above discussion, we choose a certain frequency~($\omega_{0}=1.170$) to implement the classification. However, as shown in Fig.~\ref{fig5}a, the topological states with pseudospin used for the classification span across the band gap. Therefore, frequency-division multiplexing (FDM)~\cite{li2023highly}, which splits a bandwidth into multiple, non-overlapping frequency bands, each carrying a separate harmonic signal, can be employed to implement parallel machine learning in our TMNN.

We then demonstrate the classification tasks in parallel using TMNNs. The chosen frequencies are $\omega_{0}=1.154$, $\omega_{0}=1.186$ and $\omega_{0}=1.217$, as indicated by the lines in different colors in Fig.~\ref{fig5}a, corresponding to tasks of Iris flower classification~(amplitude encoding), Penguin classification~(phase encoding) and Seeds classification~(hybrid encoding), respectively. The input signal can be simply constructed by summing individual harmonic signals after different encoding methods shown above. The batch gradient descent is used to minimize the total loss of three tasks by the averaged gradient calculated from the data in each batch. To provide a measure of the performance of our classification model for each classification task $\gamma$, the separated loss is calculated using $\mathcal{L}_{\gamma}=-[y_{\gamma}\ln{r_{\gamma}}+(1-y_{\gamma})\ln{(1-r_{\gamma})}]$, and the accuracy is computed by the ratio of correctly classified samples to the total number of samples in the task.

In Figs.~\ref{fig5}b-d, we show the training processes for three classification tasks in parallel, respectively. Generally, the loss decreases over epoch and the accuracy for the training dataset increases. Meanwhile, the accuracy for the testing dataset increases as well. This demonstrates the successful training for the TMNN. The difference between the training curve in parallel classification and that in separate classification shown in Fig.~\ref{fig3} is that the training curves in Figs.~\ref{fig5}b-d exhibit more zigzag shapes, making them less smooth. This indicates that the parallel classification can be challenging because achieving a shared TMNN for multiple tasks is difficult. This is especially true when separate classifications reveal conflicting configurations for trained TMNNs, such as general decreases in spring constants for Iris flower and Penguin classifications~(Figs.~\ref{fig3}b and Supplementary Figure 3c) while increase in spring constants for Seeds classification~(Supplementary Figure 4c).

In Fig.~\ref{fig5}e, we display the trained TMNN. We observe that the untrained, ordered system becomes the disordered system with varying spring constant distributions. Using the method mentioned above, we calculate the spin Bott index, where inner networks have an index $1$ and outer networks have an index $0$, implying the topological nature of the trained TMNN. Fig.~\ref{fig5}f gives how the histogram of changes in spring constants evolves over epoch. Compared with the separate classification, the trained TMNN for parallel classification has more widely-distributed spring constants to accommodate multiple tasks.

To demonstrate the inference process, we choose a set of data from Iris virginica, Gentoo penguin and Canadian seeds to form a composite signal after encoding. We then excite the TMNNs using the composite signal and collect the velocity field as a series of snapshots over time. To analyze the output signals and obtain inference results which are spatiotemporal patterns at each assigned frequency, we employ the dynamic mode decomposition~(DMD), a data-driven method to analyze spatiotemporal pattern evolving over time, which is widely used in wave dynamics~\cite{schmid2010dynamic,li2023characterization,li2023geometry}, biology~\cite{li2023data} and robotics~\cite{berger2015estimation}. This method provides the time dynamics and the corresponding DMD modes at different frequencies. In comparison, fast Fourier transform cannot evaluate the patterns in both space and time simultaneously. The details of this method are shown in Supplementary Information. Next, we use the patterns of DMD modes to check the correctness of classification results.

Fig.~\ref{fig5}g gives the DMD spectrum, showing the relation between the amplitudes of DMD modes and corresponding frequencies to represent the response of the entire network. It is evident that there is a region with large mode amplitudes corresponding to our excitation frequencies, indicating the dominant modes. Figs.~\ref{fig5}h-j exhibit the amplitude of DMD modes corresponding to the assigned frequencies to show the classification results. All three cases show the well-confined DMD modes along the interface, indicating the topological states. At the assigned frequency for each classification task, the DMD modes are mainly localized along the right-sided interface, right-sided interface and left-sided interface, respectively, to indicate the inference results of Iris virginica, Gentoo penguin and Canadian seeds, which agrees with the input signal.

\section{Conclusion}\label{sec3}
In conclusion, we have discovered the capability of TMNNs for performing classification towards machine learning. We have showcased that untrained TMNNs inherently cluster data, demonstrating their computational potential using pseudospin waves. For  supervised learning, we derive the in-situ backpropagation algorithm to obtain the gradient of the loss function, thereby reducing the computation load on digital computers, which are compatible with local rules in physical learning. By our training method, we demonstrate the successful training of TMNNs for classification tasks involving Iris flower, Penguin and Seeds, through amplitude encoding, phase encoding and hybrid encoding, respectively, all of which reach high classification accuracy. Moreover, we show the topologically protected robustness of TMNNs under damage by pruning bonds, where damage to most bonds barely affect the classification accuracy. Furthermore, parallel classification is demonstrated using FDM, assigning different tasks to different frequencies. This parallel computing capability allows TMNNs to perform multiple classification tasks simultaneously with a composite input, producing inference results corresponding to the assigned frequencies.

Notably, in situ backpropagation has been proposed for MNNs in statics recently~\cite{li2024training}. Our work expands this concept to the harmonic wave dynamics, filling a significant gap in the study of MNNs. By demonstrating this ubiquitous technique within mechanical dynamical systems as a physical implementation, we reveal the promising capabilities of TMNNs to reduce the cost of machine learning. The successful implementation of various tasks using TMNNs has wide-ranging implications, bridging the fields of mechanics and machine learning, and paving the way for offering efficient and resilient solutions for future machine learning hardware.

It is important to note that all key ingredients for training TMNNs and leveraging them to perform classification tasks experimentally have been  explored, facilitating further experimental investigation of TMNNs. For instance, in the stage of training TMNNs, it is necessary to measure modal displacement and modal elongation. These measurements have been demonstrated in numerous experimental efforts in the field of topological metamaterials using tools such as laser vibrometers and accelerometers~\cite{li2018observation,chaunsali2018experimental,chaunsali2018subwavelength}. Additionally, after obtaining the gradient from in situ backpropagation, the update of spring constants according to gradient descent can be conducted through tunable bars~\cite{lee2022mechanical} or stimuli-active materials, such as phase-changing materials~\cite{zappetti2020phase} and phototunable materials~\cite{stowers2015dynamic}. These materials, whose properties can be programmable in situ by external fields, hold promise as potential candidates.

In addition, several challenges remain that need to be addressed in future research. our in situ backpropagation method is conducted in the frequency domain, thereby restricted to harmonic input where the data is encoded by the amplitude and phase of harmonic waves. As a result, exploring in situ backpropagation in the time domain could be useful, as it would allow the exploitation of arbitrary waveform to encode information. Another constraint arises from the linear nature of current TMNNs, which leads to applications focused on linear classifiers. In this context, the boundaries for multiple classes are linear in the feature space. Therefore, exploring the nonlinear regime of TMNNs and leveraging nonlinear topological phenomena~\cite{snee2019edge,yang2024hall,Zhou2020,Sun2021,xiu2022topological,nunez2023fractional,xiu2023synthetically} present opportunities for both theoretical advancements in training methods and practical applications, which is a promising avenue worth exploring in future studies.

\section*{Methods}\label{sec4}
\subsection*{Numerical simulations}
The simulations of the response of TMNNs to applied harmonic forces are conducted using spring elements. Training of the TMNNs is conducted by the in situ backpropagation derived from the adjoint method, as detailed in the main text. The untrained TMNN is regarded as the initial configurations in the training process. The gradient obtained from this process is utilized to update the spring constant, using the Adam optimization algorithm. This process iterates until convergence, achieving a trained TMNN. To avoid the Anderson localization, the variation of spring constants of bonds is restricted in a range from $-0.05$ to $0.05$ and that of interfacial bonds remain unchanged, which is realized by defining a modified Sigmoid function~\cite{li2024training}. The learning rate $\alpha$ for the Iris flower classification, Penguin classification, Seeds classification and parallel classification demonstrated in the main text are $0.05$, $0.01$, $0.03$ and $0.01$, respectively. The decay rate for momentum $\beta_{1}$ and the decay rate for squared gradients $\beta_{2}$ are kept to be $0.9$ and $0.999$, respectively. Note that to make the dataset balance in the parallel classification, we use $100$ sets of data from each dataset to form the dataset instead of using all of them. For the inference process in TMNNs, we employ the Runge-Kutta method to obtain the time-dependent response under the excitation of harmonic forces encoded by the data.

\backmatter

\bmhead{Supplementary information}
See the attached Supplementary Information.

\bmhead{Acknowledgments}
The authors thank the support from the Office of Naval Research~(MURI N00014-20-1-2479) and National Science Foundation Center for Complex Particle Systems~(Award \#2243104). S.L. would like to thank Dr. Xiaotian Shi for the discussion on spin Bott index and Dr. Wenting Cheng for the insightful discussion.

\clearpage
\bibliography{sn-bibliography}

\end{document}